\documentclass[twoside,fleqn]{article}
\usepackage{epsfig}
\usepackage{espcrc2}

\title{The non-zero baryon number formulation of QCD}

\author{ O. Kaczmarek with J. Engels, F. Karsch, E. Laermann
\thanks{The work has been supported by the TMR network ERBFMRX-CT-970122
and the DFG under grant Ka 1198/4-1 .}
\\
\vskip 6pt
Fakult\"at f\"ur Physik, Universit\"at Bielefeld, D-33615 Bielefeld, Germany
}

\begin{document}

\begin{abstract}
We discuss the non-zero baryon number formulation of QCD in the quenched
limit at finite temperature. This describes the thermodynamics of gluons in the
background of static quark sources.
Although a sign problem remains in this theory,
our simulation results show that it can be handled quite well numerically. The
transition region gets shifted to smaller temperatures and the transition
region broadens with increasing baryon number. Although the action is in our
formulation explicitly $Z(3)$ symmetric the Polyakov loop expectation value
becomes non-zero already in the low temperature phase and
the heavy quark potential gets
screened at non-vanishing number density already this phase.
\vspace*{-0.4cm}
\end{abstract}
\maketitle
\section{Introduction}
An important aim of lattice QCD is the understanding of the QCD phase diagram
and its dependence on the temperature $T$ and the baryon density
$n_B$. Especially the region of non-zero density is important, as it describes
the behaviour of dense matter created in heavy ion collisions and plays an
important role in the cosmological context. Due to the well known problem of
the complex fermion determinant \cite{Barbour} when a non-zero
chemical potential $\mu$ is introduced \cite{Hasenfratz,Kogut}, only qualitative
features of the phase digram at non-zero density can be understood in terms of
models and approximations.\\
Introducing a chemical potential $\mu$ \cite{Hasenfratz} leads to the grand
canonical partition function of finite density QCD. An alternative formulation
is given in terms of the canonical partition function
at fixed non-zero baryon number \cite{Miller}.
This is
achieved by introducing an imaginary chemical
potential \cite{Miller,Roberge}
in the grand canonical partition function and
performing a Fourier integration to project onto the canonical partition function
for a given sector of fixed baryon number \cite{Miller}
\begin{eqnarray}
Z(B,T,V) = \frac{1}{2\pi} \int_0^{2\pi} \mathrm{ d}\phi e^{-iB\phi} Z(i\phi,T,V).
\end{eqnarray}
This formulation still suffers from a sign problem, but leads to a quite
natural and useful formulation of the quenched limit of QCD at non-zero baryon
number density, where the sign problem can be handled quite well \cite{kacz}.
\section{The quenched limit of QCD at non-zero density}
The static limit of QCD at non-zero chemical potential $\mu$ in the grand canonical
approach has been formulated in \cite{Stam} and \cite{Blum}.
The numerical results indicate
that in this case the first
order deconfinement transition of the $SU(3)$ gauge theory turns into a
crossover for arbitrarily small, non-zero values of the chemical
potential \cite{Blum}. The quenched partition function in the canonical
approach can be written as
\begin{eqnarray}
Z(B,T,V) = \int \prod_{x,\nu} \mathrm{ d}U_{x,\nu} \hat f_B e^{-S_G}
\end{eqnarray}
where the constraint on the baryon number is encoded in the function $\hat
f_B$ which is a function of Polyakov loops, $B$ counts the number of quarks,
i.e. $B/3$ is the baryon number. For $B=3$ and one flavour of Wilson fermions
$\hat f_B$ is, for instance, given by
\begin{eqnarray}
f_{B=3} &=& (2\kappa)^{3N_\tau} ( V^3 \frac{4}{3} [ L_{1,0}]^3 \nonumber\\
&& + V^2 (8[ L_{1,0}][ L_{2,0} ] - 4[ L_{1,0}
][ L_{0,1} ] )\nonumber\\
&& + V (12 + \frac{2}{3} [ L_{3,0} ] - 2 [ L_{1,1} ] ) )
\end{eqnarray}
\hspace*{0.2cm} with $[ L_{i,j} ] = V^{-1} \sum_{\vec x} ({\rm Tr}
L_{\vec x} )^i ({\rm Tr} L^2_{\vec x})^j$.\medskip\\
\begin{figure}[t]
\begin{center}
\epsfig{file=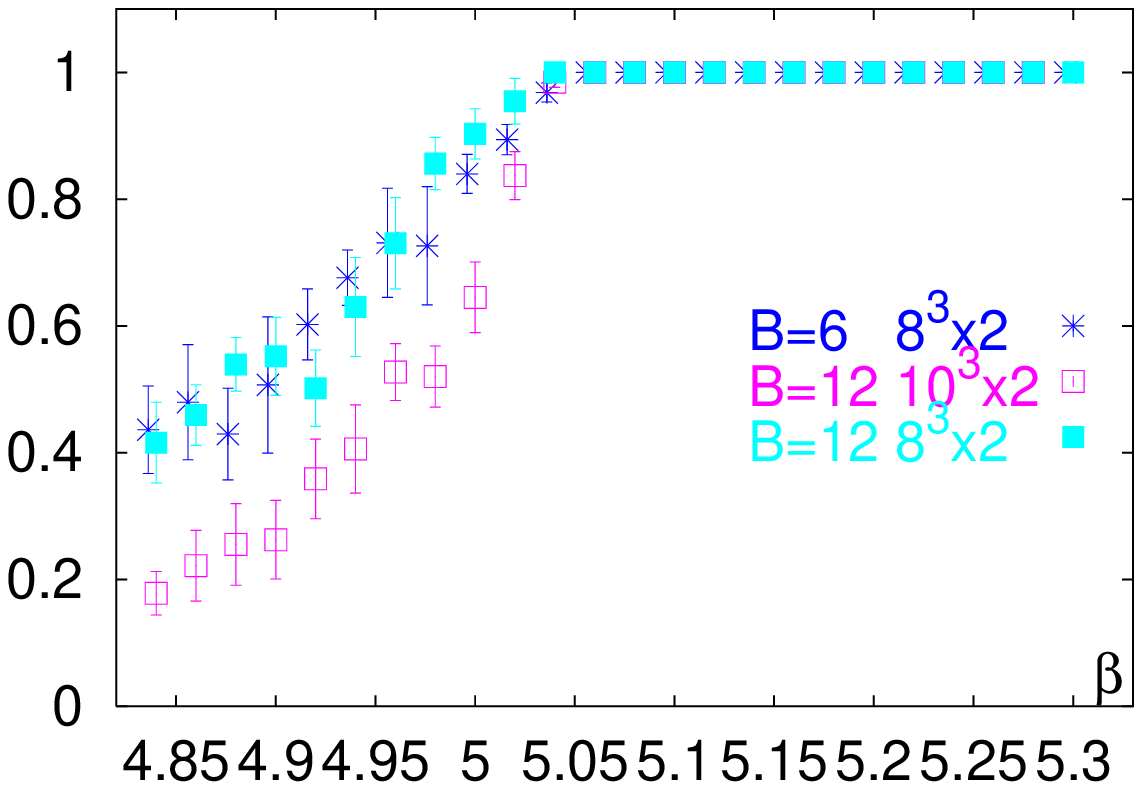, width=7cm}
\epsfig{file=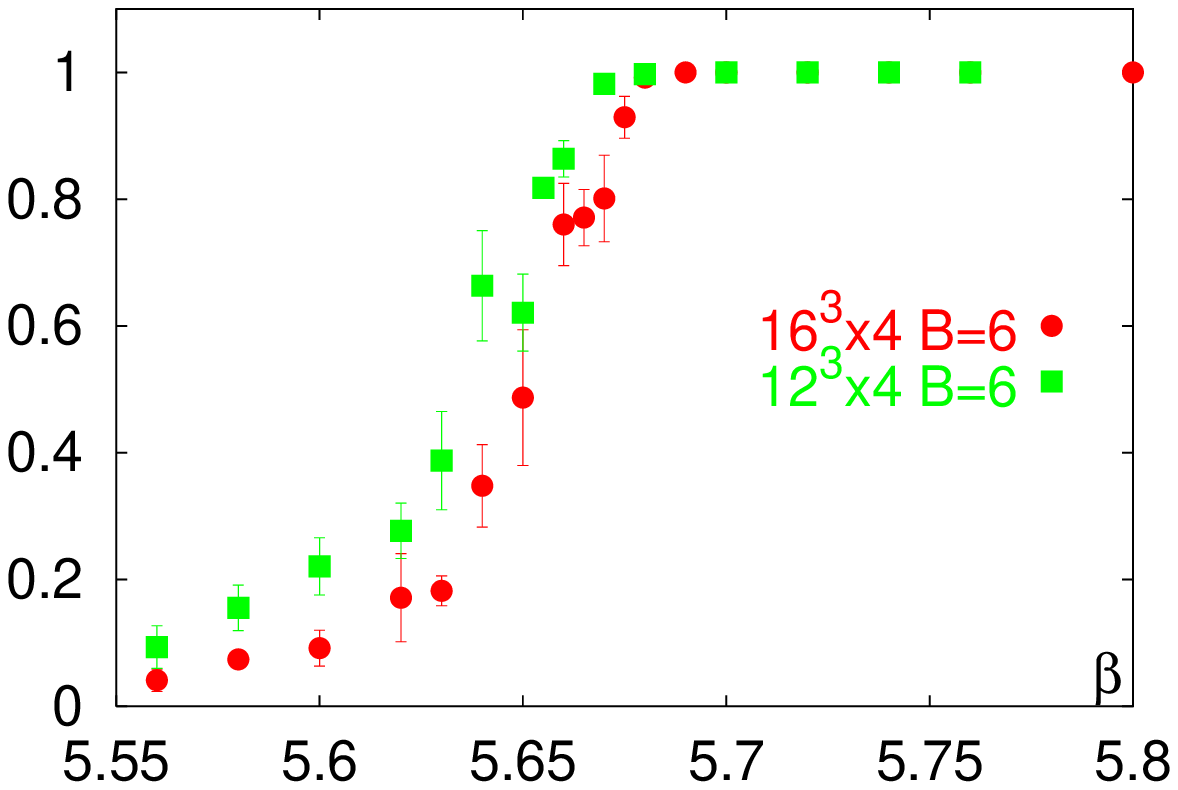, width=7cm}
\vspace*{-1cm}
\caption{$\langle \mathrm{sgn}(\mathrm{ Re} \hat f_B)\rangle$ for
  $B=6$ and 12 calculated on lattices of size $N_\sigma^3\times N_\tau$.}
\vspace*{-0.4cm}
\end{center}
\end{figure}
For a more detailed description of $\hat f_B$ see \cite{kacz}. $S_G$ is the
gluonic action, which is $Z(3)$ symmetric. The partition function $Z(B,T,V)$,
is non-zero only if $B$ is a multiple of 3, because $\hat f_B$ is invariant
under $Z(3)$ transformations only if $B$ is a multiple of 3. In general it
changes by a factor $e^{2\pi i B/3}$ under a global $Z(3)$ transformation of
time-like link variables.
\begin{figure}[t]
\begin{center}
\epsfig{file=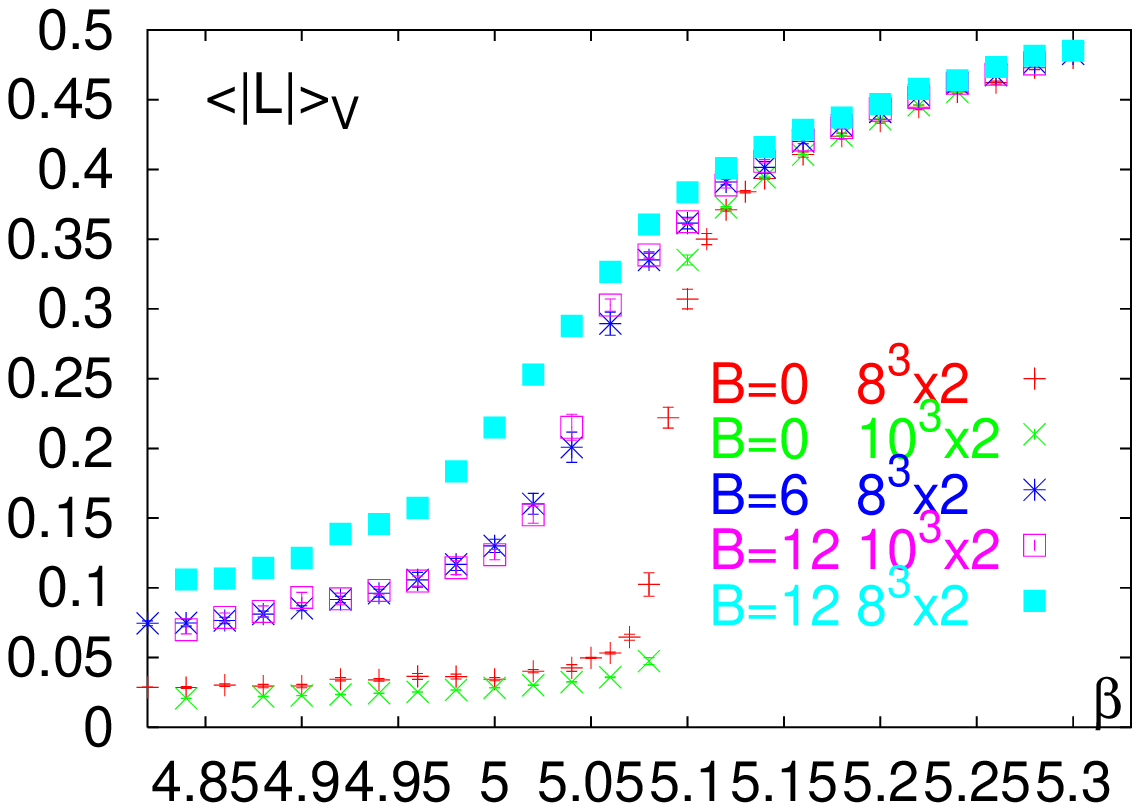, width=7cm}
\epsfig{file=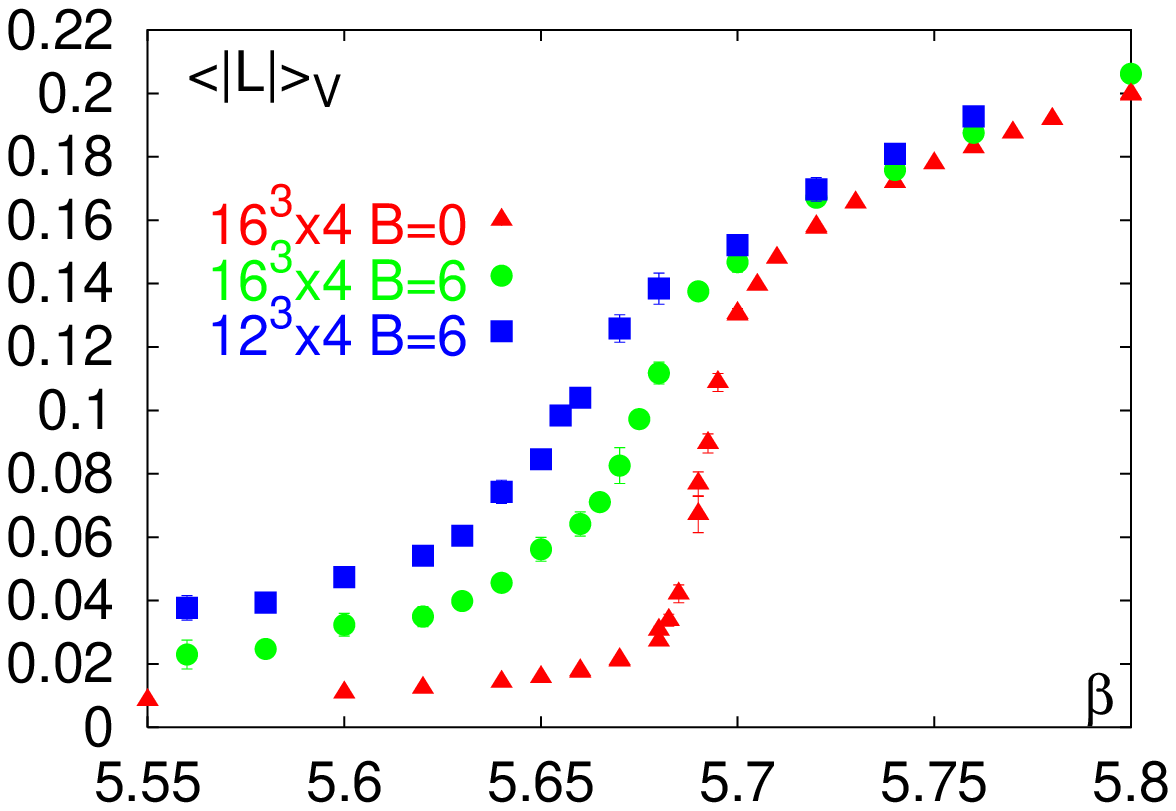, width=7cm}
\vspace*{-1cm}
\caption{Polyakov loop expectation value $\langle \vert L \vert \rangle_V$
for different values of $B$ and
lattices of size $N_\sigma^3\times N_\tau$.}
\vspace*{-0.4cm}
\end{center}
\end{figure}
\begin{figure}[t]
\begin{center}
\epsfig{file=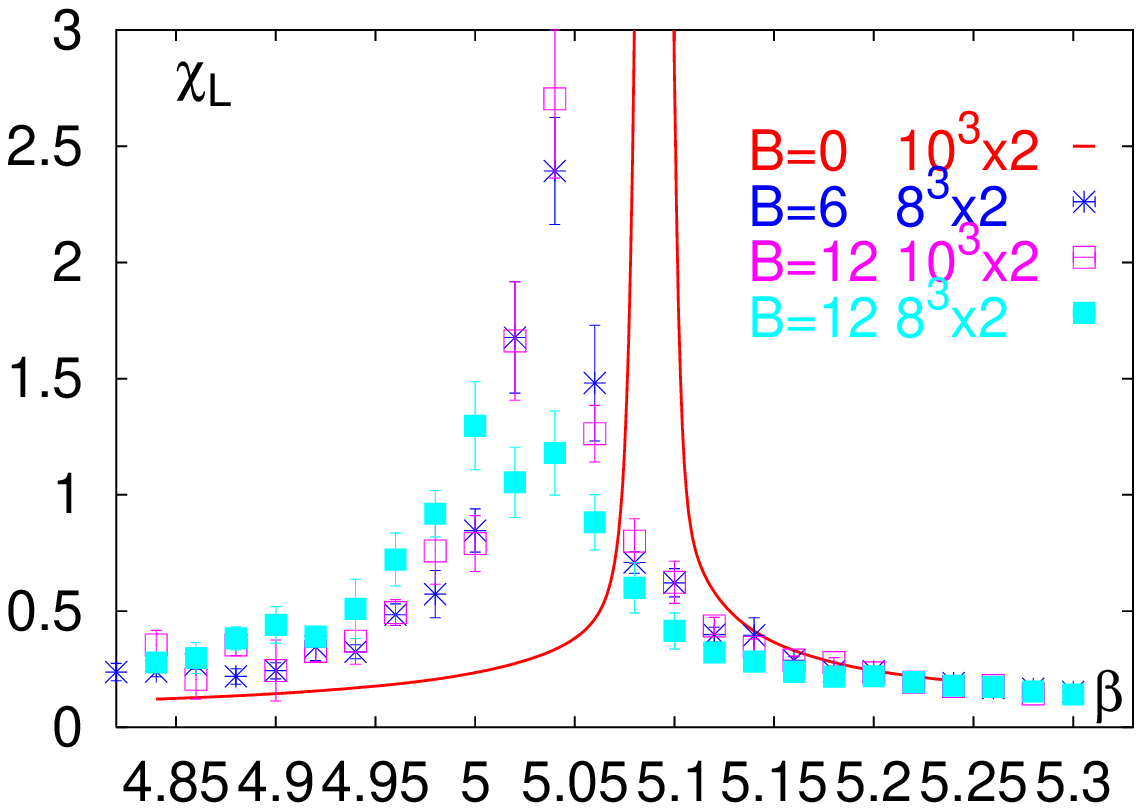, width=7cm}
\epsfig{file=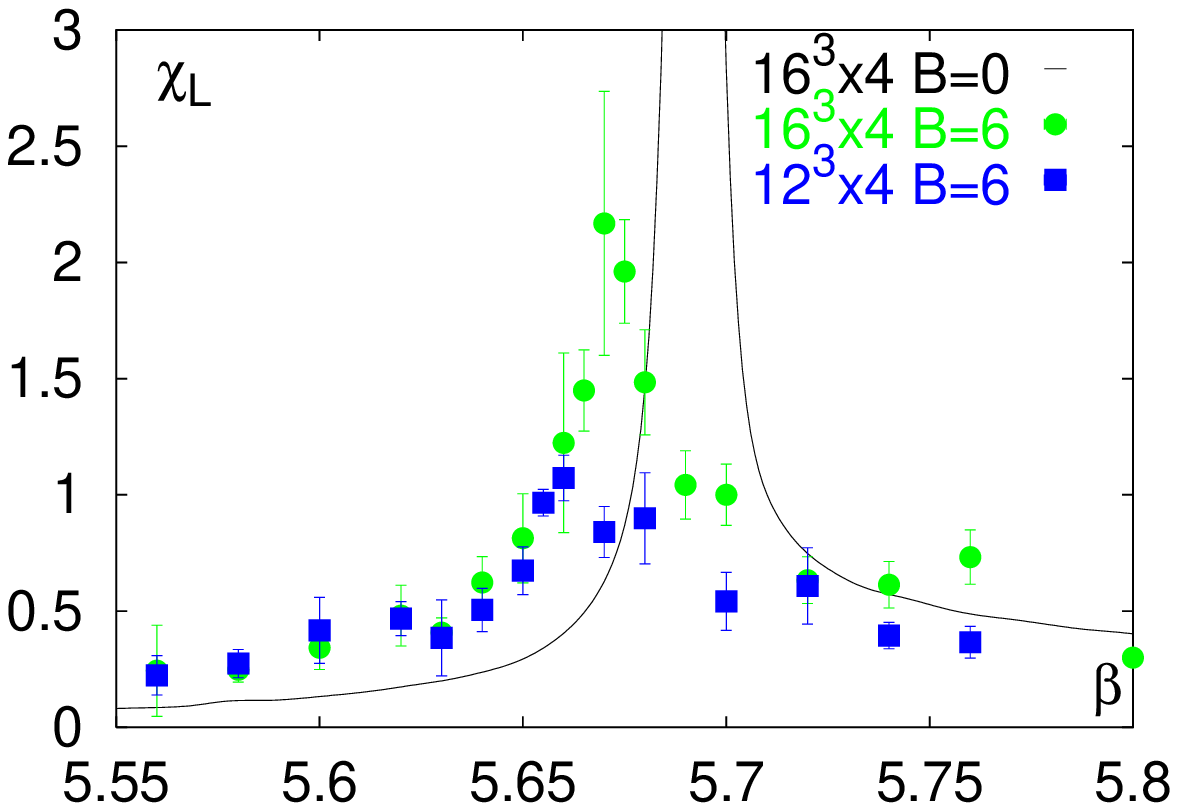, width=7cm}
\vspace*{-1cm}
\caption{Polyakov loop
  susceptibility $\chi_L$ for different values of $B$ and
  lattices of size $N_\sigma^3\times N_\tau$.}
\vspace*{-0.4cm}
\end{center}
\end{figure}
$\hat f_B$ is still a complex function, but upon integration over the gauge
fields the imaginary part of the partition function vanishes. The remaining sign problem can be
handled by using the absolute value of $\mathrm{Re} \hat f_B$ and
including the sign in the calculation of observables \cite{Engels}.\\
Our simulations are performed on $N_\sigma^3 \times N_\tau$ lattices
with $N_\sigma=8,10,12,16$ and $N_\tau=2,4$ using the
standard Wilson action and one flavour of Wilson fermions with quark number values of $B=6$ and
12 at fixed $n_B/T^3=(1/3) B (N_\tau/N_\sigma)^3$.
Fig. 1 shows the average sign $\langle \mathrm{ sgn}(\mathrm{Re}\hat
f_B)\rangle$ as a function of the coupling $\beta$.
For large values of the temperature the sign is almost always positive, but also for
the smallest temperature in our analysis the sign can be well determined. It
depends on the spatial volume $N_\sigma^3$ but varies little with $B$.
\begin{figure}[t]
\begin{center}
\epsfig{file=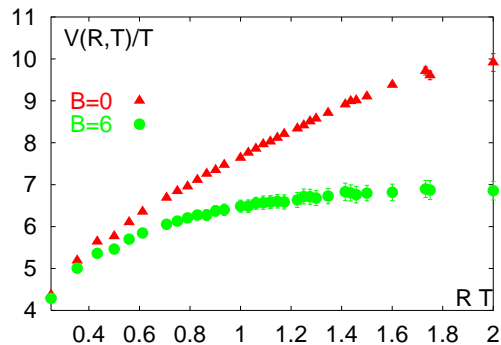, width=7cm}
\vspace*{-1cm}
\caption{Heavy quark potential for $\beta=5.62$ and $B=0$ and 6.}
\vspace*{-0.4cm}
\end{center}
\end{figure}
The Polyakov loop expectation values in Fig. 2 show a clear signal for
a first order transition for the $B=0$
case, while for all $B>0$ the transition is continuous. The transition
region is shifted towards smaller $\beta$-values and it
broadens with increasing $B$. Note that by changing the gauge coupling $\beta$ we vary the
lattice cut-off and through this also the baryon number density
continuously. The broadening of the transition region may indicate the presence
of a region of coexisting phases. The Polyakov loop susceptibility (Fig. 3)
reflects the existence of
a transition region that becomes broader with increasing $n_B$, but does
not show indications for a discontinuity.\\
The Polyakov loop expectation value becomes non-zero already in the low
temperature phase. This indicates that the heavy quark potential stays finite
at large distances. We validate this by calculating the potential using
Polyakov loop correlations (Fig. 4). For zero
baryon number it shows the
usual linearly rising behaviour for the quenched case. For $B=6$ the potential
stays finite at large distances due to the screening of
the static quark anti-quark sources by already present static quarks.
This behaviour is comparable to heavy quark potentials in full QCD \cite{Edwin}.
\section{Conclusions}
We have analyzed the quenched limit of QCD at non-zero baryon number. The sign
problem in this theory can be handled quite well
numerically. We find indications for a region of coexisting phases, which broadens with
increasing baryon number density and is shifted towards smaller
temperatures. Further analyses are needed to see if this is a signal of the
existence of a first order phase transition or a smooth crossover at non-zero density.
We also see evidence that the heavy quark potential for non-zero baryon density
stays finite for large distances already in the hadronic phase. The potential gets screened
by the static quarks that induce the non-vanishing density. This will have a
direct influence on heavy quark bound states at high density.

\end{document}